\documentclass{llncs}
\usepackage{amssymb}
\usepackage{amsmath}
\usepackage{amsfonts}
\usepackage{graphicx}

\input tcilatex
\begin{document}

\author{Zining Cao \\
\institute{
          College of Computer Science and Technology \\ Nanjing University of Aeronautics and Astronautics, Nanjing 211106, China \\
    \email{caozn@nuaa.edu.cn}}}
\title{On Generalized Performance Evaluation and Generalized Controller
Synthesis}
\maketitle

\begin{abstract}
In this paper, we propose the frameworks of generalized performance
evaluation and generalized controller synthesis. To this end, we give a true
concurrent process calculus as the model of systems, and present a
complete-lattice-monoid-valued performance evaluation language as the
performance specification of systems. We give a framework of generalized
performance evaluation based on the process calculus and the performance
evaluation language. We show that the several problems in computer science
are special cases of generalized performance evaluation. A generalized
performance evaluation algorithm is presented. Furthermore, we present a
framework of generalized controller synthesis, which is the inverse problem
of generalized performance evaluation. We show several special cases of
generalized controller synthesis in computer science, and give an outline of
generalized controller synthesis algorithm.
\end{abstract}

\section{Introduction}

Performance evaluation aims at analysing quantitative system aspects that
are related to its performance and dependability. Evaluation approaches are
measurement-based and model-based techniques. In measurement-based
techniques, controlled experiments are performed on a concrete realisation
of the system, and gathered timing information is analysed to evaluate the
measures of interest such as time-to-failure or system throughput. In
model-based performance evaluation, an abstract model of the system is
constructed that is just detailed enough to evaluate the measures of
interest with the required accuracy. Depending on modelling flexibility and
computational requirements, either analytical, numerical or simulative
techniques are used to evaluate the required measures. Model specification
and analysis techniques are key ingredients of model-based performance
evaluation methodology \cite{BDDHP11}, \cite{BHH05}, \cite{HK01}, \cite%
{JM18a}, \cite{JM18b}, and \cite{MJ10}.

Controller synthesis is a fundamental concept within control theory and
computer science. It refers to the constructing a controller for a dynamic
system to ensure it behaves in a desired manner. In essence, it is the
inverse problem of performance evaluation. While performance evaluation
focuses on evaluating the performance properties of an existing system,
synthesis is the methodology for creating the controller itself to meet
specific design objectives. The primary goal of controller synthesis is to
determine the structure and parameters of a controller that will make the
overall closed-loop system achieve desired performance specifications, such
as stability, robustness against uncertainties, and specific dynamic
responses \cite{CK98}, \cite{HSLL10}, and \cite{VSSLS01}.

In this paper, we generalize the value of performance evaluation from real
number domain to complete-lattice-monoid domain. The methodology is based on
appropriate extensions of process calculus for the description of systems
and model checking techniques for their analysis. We show that the
generalized performance evaluation can be viewed as a uniform framework of
several different problems in formal methods and computational complexity.
Moreover, we present generalized controller synthesis as the inverse problem
of generalized performance evaluation, and show that generalized controller
synthesis can be viewed as a uniform framework of some special cases in
formal methods and control theory. To give the framework of generalized
performance evaluation and generalized controller synthesis, we present a
true concurrent process calculus as the model of systems, and present a
complete-lattice-monoid-valued performance evaluation language as the
performance specification of systems. The generalized performance evaluation
algorithm and generalized controller synthesis algorithm are also given in
the case that the process space and the complete lattice are both finite.

This paper is organized as follows: Section 2 gives a true concurrent
process calculus including its syntax, operational semantics and comparison
with Turing machine and Petri net. In Section 3, we present a
complete-lattice-monoid-valued performance evaluation language. Section 4
gives a framework of generalized performance evaluation based on the true
concurrent process calculus and the complete-lattice-monoid-valued
performance evaluation language. The problems of model checking,
bisimulation checking and computational complexity are all special cases of
generalized performance evaluation. Section 4 also presents a generalized
performance evaluation algorithm. Section 5 gives a framework of generalized
controller synthesis. The problems of program synthesis, counterexample
generation, bisimilar process generation and controller synthesis are all
special cases of generalized controller synthesis. An outline of generalized
controller synthesis algorithm is given in Section 5. The paper is concluded
in Section 6.

\section{True Concurrent Process Calculus $PROC$}

Process calculus is a family of formal languages used to model and reason
about concurrent, communicating, and distributed systems. It provides a
mathematical framework for describing the behavior of interacting processes.
In this section, we propose a true concurrent process calculus $PROC$ as the
model of systems for performance evaluation. We also give some bisimulations
of $PROC$.

\subsection{Syntax of True Concurrent Process Calculus $PROC$}

In the following, we give the syntax of $PROC.$

\textbf{Definition 1.} Let $Const$ be a finite set of process constants. The
classes of process expressions called $PROC$ are defined by the following
abstract syntax:

$P::=$ $\epsilon $ $|$ $C$ $|$ $P_{1};P_{2}$ $|$ $P_{1}+P_{2}$ $|$ $%
P_{1}\otimes _{L}P_{2}$ $|$ $(\nu a)P,$ where $\epsilon $ is the empty
process, $C$ ranges over $Const$, the operator $;$ stands for a sequential
composition, $+$ stands for a nondeterministic composition, $AtomAct$ is the
set of atomic action$,$ $L\subseteq AtomAct,$ $\otimes _{L}$ stands for a
parallel composition, $a\in AtomAct,$ $(\nu a)$ is a restriction. In the
following, we abbreviate $(\nu a_{1})...(\nu a_{n})P$ as $(\nu L)P$ where $%
L=\{a_{1},...,a_{n}\}.$

We do not distinguish between process expressions related by a structural
congruence, which is the smallest congruence over process expressions such
that the following laws hold: $;$ is associative, $+$ and $\otimes _{L}$ are
associative and commutative, and $\epsilon $ is a unit for $;$, $+$ and $%
\otimes _{L}$.

\subsection{Labelled Transition System of True Concurrent Process Calculus $%
PROC$}

We give the operational semantics of $PROC$ in the form of labelled
transition system.

\textbf{Definition 2.} A set of basic transition rules is the finite set $%
\Delta \subseteq (Const\backslash \{\epsilon \})\times Act\times PROC,$
where $Act$ is the set of all finite multisets of $AtomAct$.

In the following, we adopt $\{\{x_{1},...,x_{n}\}\}$ to denote finite
multiset with elements $x_{1},...,x_{n}$.

A labelled transition system of the class $PROC$ (modulo the structural
congruence) is a transition relation which is the least relation satisfying
the following rules (recall that $+$ and $\otimes _{L}$ are commutative):

\begin{center}
$T-CON:\dfrac{(C,\alpha ,P)\in \Delta }{C\overset{\alpha }{\longrightarrow }P%
}$

$T-SEQ:\dfrac{P\overset{A}{\longrightarrow }P^{\prime }}{P;Q\overset{A}{%
\longrightarrow }P^{\prime };Q}$

$T-CHO:\dfrac{P\overset{A}{\longrightarrow }P^{\prime }}{P+Q\overset{A}{%
\longrightarrow }P^{\prime }}$

$T-COM:\dfrac{P_{1}\overset{A}{\longrightarrow }P_{1}^{\prime },P_{2}\overset%
{B}{\longrightarrow }P_{2}^{\prime }}{P_{1}\otimes _{L}P_{2}\overset{%
(A\uplus B)-L}{\longrightarrow }P_{1}^{\prime }\otimes _{L}P_{2}^{\prime }}$
where $A\cap L=B\cap L$.

$T-RES:\dfrac{P\overset{A}{\longrightarrow }P^{\prime }}{(\nu a)P\overset{%
A(\tau /a)}{\longrightarrow }(\nu a)P^{\prime }}$ where $A(\tau /a)$ denotes
the set by replacing $a$ by $\tau $ in the set $A$.

Table 1: Operational semantics of true concurrent process calculus
\end{center}

Similar to process calculus, in each process of the form $(\nu a)P$ the
occurrence of $a$ is bound within the scope of $P$. An occurrence of $a$ in $%
P$ is said to be free iff it does not lie within the scope of a bound
occurrence of $a$. The set of names occurring free in $P$ is denoted $fn(P)$%
. An occurrence of a name in $P$ is said to be bound if it is not free, we
write the set of bound names as $bn(P)$. $n(P)$ denotes the set of names of $%
P$, i.e., $n(P)=fn(P)\cup bn(P)$.

$A\uplus B$ denotes the union set of multiset $A$ and $B,$ for example, $%
\{\{a,a,b\}\}\uplus \{\{a,b,b,c\}\}=\{\{a,a,a,b,b,b,c\}\}$. $A-B$ denotes
the difference set of multiset $A$ and $B,$ for example, $%
\{\{a,a,b\}\}-\{\{a,b,b,c\}\}=\{\{a\}\}$. If $L$ is a set, $L$ is viewed as
a multiset in $A-L.$ $A\cap L$ denotes the intersection set of multiset $A$
and $L,$ for example, $\{\{a,a,b,b\}\}\cap \{\{a,b,b,b,c\}\}=\{a,b,b\}.$ If $%
L$ is a set, $L$ is viewed as a multiset in $A\cap L.$ $A(\tau /a)$ denotes
the set by replacing $a$ by $\tau $ in the multiset $A,$ for example, $%
\{\{a,a,a,b,b,b,c\}\}(\tau /a)=\{\{\tau ,\tau ,\tau ,b,b,b,c\}\}$.

Structural congruence: $P+Q\equiv Q+P;$ $P\otimes _{L}Q\equiv Q\otimes
_{L}P; $ $(P;Q);R\equiv P;(Q;R);$ $(P+Q)+R\equiv P+(Q+R);$ $(P\otimes
_{L}Q)\otimes _{L}R\equiv P\otimes _{L}(Q\otimes _{L}R);$ $P;\epsilon \equiv
P;$ $\epsilon ;P\equiv P;$ $P+\epsilon \equiv P;$ $P\otimes _{L}\epsilon
\equiv P;$ $(\nu a)\epsilon \equiv \epsilon ;$ $(\nu a)(\nu b)P\equiv (\nu
b)(\nu a)P;$ $(\nu a)(P+Q)\equiv P+(\nu a)Q\ $if $a\notin fn(P);$ $(\nu
a)(P\otimes _{L}Q)\equiv P\otimes _{L}(\nu a)Q\ $if $a\notin fn(P).$

By the labelled transition system of $PROC,$ it is easy to see that any $%
PROC $ process is image-finite.

\textbf{Lemma 1}. Every $PROC$ process $P$ is image-finite, i.e., the set $%
\{Q$ $|$ $P\overset{a}{\longrightarrow }Q\}$ is finite for any multiset
action $a$.

\subsection{Bisimulation of $PROC$}

Now we propose some bisimulations for process calculus $PROC$. Bisimulation
is a fundamental concept in theoretical computer science, particularly in
the study of concurrent and reactive systems. It is a binary relation
between state transition systems, where two systems are considered bisimilar
if they can mutually simulate each other's behavior step by step, making
them indistinguishable to an external observer.

\textbf{Definition 3.} A symmetric relation $R\subseteq PROC\times PROC$ is
a strong bisimulation if whenever $(P,Q)\in R$, $P\overset{A}{%
\longrightarrow }P^{\prime }$ implies that there exists $Q^{\prime }$ such
that $Q\overset{A}{\longrightarrow }Q^{\prime }$ and $(P^{\prime },Q^{\prime
})\in R$.

We write $P\sim Q$ if there is a strong bisimulation $R$ such that $(P,Q)\in
R$.

\textbf{Definition 4.} A symmetric relation $R\subseteq PROC\times PROC$ is
a weak bisimulation if whenever $(P,Q)\in R$, $P\overset{A-\{\{\tau
,...,\tau \}\}}{\longrightarrow }P^{\prime }$ implies that there exists $%
Q^{\prime }$ such that $Q\overset{\varepsilon }{\Longrightarrow }\overset{%
A-\{\{\tau ,...,\tau \}\}}{\longrightarrow }\overset{\varepsilon }{%
\Longrightarrow }Q^{\prime }$ and $(P^{\prime },Q^{\prime })\in R,$ where $%
\overset{\varepsilon }{\Longrightarrow }$ is $(\overset{\{\{\tau ,...,\tau
\}\}}{\longrightarrow })^{\ast },\{\{\tau ,...,\tau \}\}$ means $\varnothing 
$ or the set of $n$ times of $\tau ,n\in \mathbf{N}$.

We write $P\approx Q$ if there is a weak bisimulation $R$ such that $%
(P,Q)\in R$.

\textbf{Definition 5.} The strong bisimulation approximants $\sim _{\kappa
}\subseteq PROC\times PROC$, for all ordinals $\kappa \in O$, are defined as
follows:

(1) $P\sim _{0}Q$ for all process $P$ and $Q$.

(2) $P\sim _{\kappa +1}Q$ iff

(a) for each transition $P\overset{A}{\longrightarrow }P^{\prime }$ there is
a transition $Q\overset{A}{\longrightarrow }Q^{\prime }$ such that $%
P^{\prime }\sim _{\kappa }Q^{\prime }$; and

(b) for each transition $Q\overset{A}{\longrightarrow }Q^{\prime }$ there is
a transition $P\overset{A}{\longrightarrow }P^{\prime }$ such that $%
P^{\prime }\sim _{\kappa }Q^{\prime }$.

(3) For all limit ordinals $\lambda $, $P\sim _{\lambda }Q$ iff $P\sim
_{\kappa }Q$ for all $\kappa <\lambda $.

\textbf{Definition 6.} The weak bisimulation approximants $\approx _{\kappa
}\subseteq PROC\times PROC$, for all ordinals $\kappa \in O$, are defined as
follows:

(1) $P\approx _{0}Q$ for all process $P$ and $Q$.

(2) $P\approx _{\kappa +1}Q$ iff

(a) for each transition $P\overset{A-\{\{\tau ,...,\tau \}\}}{%
\longrightarrow }P^{\prime }$ there is a transition $Q\overset{\varepsilon }{%
\Longrightarrow }\overset{A-\{\{\tau ,...,\tau \}\}}{\longrightarrow }%
\overset{\varepsilon }{\Longrightarrow }Q^{\prime }$ such that $P^{\prime
}\approx _{\kappa }Q^{\prime }$; and

(b) for each transition $Q\overset{A-\{\{\tau ,...,\tau \}\}}{%
\longrightarrow }Q^{\prime }$ there is a transition $P\overset{\varepsilon }{%
\Longrightarrow }\overset{A-\{\{\tau ,...,\tau \}\}}{\longrightarrow }%
\overset{\varepsilon }{\Longrightarrow }P^{\prime }$ such that $P^{\prime
}\approx _{\kappa }Q^{\prime }$.

(3) For all limit ordinals $\lambda $, $P\approx _{\lambda }Q$ iff $P\approx
_{\kappa }Q$ for all $\kappa <\lambda $.

The following conjecture is the generalization of conjecture about the
equivalence between weak bisimulation $\approx $ and weak bisimulation
approximants $\approx _{\omega ^{\omega }}$ for $BPA$ processes \cite{S99}.

\textbf{Conjecture 1.} For any $P$ and $Q$, $P\approx Q\Leftrightarrow
P\approx _{\omega ^{\omega }}Q.$

\subsection{Comparison with Turing Machine and Petri Net}

In the following, we show that Turing machine and Petri net can be simulated
by true concurrent process calculus.

A Turing machine is a fundamental model of computation that defines the
theoretical limits of what can be calculated. A Turing machine consists of
three main parts: an infinite tape which is a long strip divided into
discrete cells holding a symbol from a finite alphabet; a read/write head
that can scan the cell, read the symbol, write a new symbol, or erase it; a
state register and instruction table which dictates actions based on the
current state and the symbol read from the tape.

\textbf{Definition 7.} a one-tape Turing machine can be formally defined as
a 7-tuple $M=\langle Q,\Gamma ,b,\Sigma ,\delta ,q_{0},F\rangle $ where

(1) $Q$ is a finite, non-empty set of states.

(2) $\Gamma $ is a finite, non-empty set of tape alphabet symbols.

(3) $b\in \Gamma $ is the blank symbol (the only symbol allowed to occur on
the tape infinitely often at any step during the computation).

(4) $\Sigma \subseteq \Gamma \setminus \{b\}$ is the set of input symbols,
that is, the set of symbols allowed to appear in the initial tape contents.

(5) $q_{0}\in Q$ is the initial state.

(6) $F\subseteq Q$ is the set of final states or accepting states. The
initial tape contents is said to be accepted by $M$ if it eventually halts
in a state from $F$.

(7) $\delta :(Q\setminus F)\times \Gamma \rightarrow Q\times \Gamma \times
\{L,R\}$ is a partial function called the transition function, where $L$ is
left shift, $R$ is right shift. If $\delta $ is not defined on the current
state and the current tape symbol, then the machine halts.

Intuitively, the following gives a transformation from a Turing machine to
the parallel composition of three $BPA$ processes.

\textbf{Definition 8.} The transformation of Turing machine $M=\langle
Q,\Gamma ,b,\Sigma ,\delta ,q_{0},F\rangle $ with configuration $C_{M}$ to
true concurrent process calculus system $Trans_{T}(C_{M},M)=P\equiv
(Q_{P}=(\nu L)(P_{1}\otimes _{L}P_{2}\otimes _{L}P_{3}),R_{P})$ as follows
(where $R_{P}$ is the set of basic transition rules)$:$

Suppose a Turing machine $M=\langle Q,\Gamma ,b,\Sigma ,\delta
,q_{0},F\rangle $, $C_{M}$ is a configuration of $M$, $C_{M}=\overline{%
\underline{...|b|C_{1}|C_{2}\uparrow q|C_{3}|b|...}}=\overline{\underline{%
...|b|\alpha _{1}|...|\alpha _{m}|\beta \uparrow q|\gamma _{1}|...|\gamma
_{n}|b|...}},$ where $q$ is the current state, $C_{2}$ $=\beta $ is the tape
alphabet symbol pointed by read/write head, $C_{1}$ $=\alpha _{1}...\alpha
_{m}$ and $C_{3}=\gamma _{1}...\gamma _{n}$ are strings of tape alphabet
symbol, and all other cells hold symbol $b$. We construct processes $(\nu
L)(P_{1}\otimes _{L}P_{2}\otimes _{L}P_{3})$ corresponding to configuration $%
C_{M}$ as follows: $P_{1}=(\alpha _{m})_{l};...;(\alpha _{1})_{l};(b)_{l},$ $%
P_{2}=\beta \uparrow q,$ $P_{3}=(\gamma _{1})_{r};...;(\gamma
_{n})_{r};(b)_{r}.$ The set of basic transition rules $\Delta _{P}$ is
constructed as follows: (1) $\beta \uparrow q\overset{\{\{(\alpha _{m},\beta
,\gamma _{1},q,L,\beta ^{\prime },q^{\prime })\}\}}{\longrightarrow }\alpha
_{m}\uparrow q^{\prime },$ $(\alpha _{m})_{l}\overset{\{\{(\alpha _{m},\beta
,\gamma _{1},q,L,\beta ^{\prime },q^{\prime })\}\}}{\longrightarrow }%
\epsilon $ ($\epsilon $ is the empty process)$,$ $(\gamma _{1})_{r}\overset{%
\{\{(\alpha _{m},\beta ,\gamma _{1},q,L,\beta ^{\prime },q^{\prime })\}\}}{%
\longrightarrow }(\beta ^{\prime })_{r};(\gamma _{1})_{r},$ if $\delta
(q,\beta )=(q^{\prime },\beta ^{\prime },L)$ and $\alpha _{m}\neq b;$ (2) $%
\beta \uparrow q\overset{\{\{(\alpha _{m},\beta ,\gamma _{1},q,L,\beta
^{\prime },q^{\prime })\}\}}{\longrightarrow }b\uparrow q^{\prime },$ $%
(b)_{l}\overset{\{\{(\alpha _{m},\beta ,\gamma _{1},q,L,\beta ^{\prime
},q^{\prime })\}\}}{\longrightarrow }(b)_{l},$ $(\gamma _{1})_{r}\overset{%
\{\{(\alpha _{m},\beta ,\gamma _{1},q,L,\beta ^{\prime },q^{\prime })\}\}}{%
\longrightarrow }(\beta ^{\prime })_{r};(\gamma _{1})_{r},$ if $\delta
(q,\beta )=(q^{\prime },\beta ^{\prime },L)$ and $\alpha _{m}=b;$ (3) $\beta
\uparrow q\overset{\{\{(\alpha _{m},\beta ,\gamma _{1},q,R,\beta ^{\prime
},q^{\prime })\}\}}{\longrightarrow }\gamma _{1}\uparrow q^{\prime },$ $%
(\alpha _{m})_{l}\overset{\{\{(\alpha _{m},\beta ,\gamma _{1},q,R,\beta
^{\prime },q^{\prime })\}\}}{\longrightarrow }(\beta ^{\prime })_{l};(\alpha
_{m})_{l},$ $(\gamma _{1})_{r}\overset{\{\{(\alpha _{m},\beta ,\gamma
_{1},q,R,\beta ^{\prime },q^{\prime })\}\}}{\longrightarrow }\epsilon ,$ if $%
\delta (q,\beta )=(q^{\prime },\beta ^{\prime },R)$ and $\gamma _{1}\neq b;$
(4) $\beta \uparrow q\overset{\{\{(\alpha _{m},\beta ,\gamma _{1},q,R,\beta
^{\prime },q^{\prime })\}\}}{\longrightarrow }b\uparrow q^{\prime },(\alpha
_{m})_{l}\overset{\{\{(\alpha _{m},\beta ,\gamma _{1},q,R,\beta ^{\prime
},q^{\prime })\}\}}{\longrightarrow }(\beta ^{\prime })_{l};(\alpha
_{m})_{l},$ $(b)_{r}\overset{\{\{(\alpha _{m},\beta ,\gamma _{1},q,R,\beta
^{\prime },q^{\prime })\}\}}{\longrightarrow }(b)_{r},$ if $\delta (q,\beta
)=(q^{\prime },\beta ^{\prime },R)$ and $\gamma _{1}=b.$

We have $(\nu L)(P_{1}\otimes _{L}P_{2}\otimes _{L}P_{3})\overset{\{\{\tau
\}\}}{\longrightarrow }(\nu L)(P_{1}^{\prime }\otimes _{L}P_{2}^{\prime
}\otimes _{L}P_{3}^{\prime })$ where $L=\Gamma \times \Gamma \times \Gamma
\times Q\times \{L,R\}\times Q,$ if configuration $C_{M}$ transforms to
configuration $C_{M}^{\prime }$ by one step, and $(\nu L)(P_{1}\otimes
_{L}P_{2}\otimes _{L}P_{3})$ correspond to configuration $C_{M},$ $(\nu
L)(P_{1}^{\prime }\otimes _{L}P_{2}^{\prime }\otimes _{L}P_{3}^{\prime })$
correspond to configuration $C_{M}^{\prime }$.

The definition of simulation $\preceq _{TM}$ can be given as follows.

\textbf{Definition 9.} For Turing machine $M=\langle Q,\Gamma ,b,\Sigma
,\delta ,q_{0},F\rangle $ with configuration $C_{M}$ and true concurrent
process calculus system $P=(Q_{P},R_{P})$ where $R_{P}$ is the set of basic
transition rules$,$ a relation $R$ is a simulation if whenever $%
((C_{M},M),(Q_{P},R_{P}))\in R$, $C_{M}\longrightarrow _{M}C_{M}^{\prime }$
implies that there exists $Q_{P}^{\prime }$ such that $Q_{P}\overset{%
\{\{\tau \}\}}{\longrightarrow }_{P}Q_{P}^{\prime }$ and $((C_{M}^{\prime
},M),(Q_{P}^{\prime },R_{P}))\in R,$ where $C_{M}\longrightarrow
_{M}C_{M}^{\prime }$ denotes that Turing machine $M\ $transforms from
configuration $C_{M}$ to configuration $C_{M}^{\prime }$ by one step, and $%
Q_{P}\overset{\{\{\tau \}\}}{\longrightarrow }_{P}Q_{P}^{\prime }$ denotes
that process $Q_{P}$ performs action $\{\{\tau \}\}$ to process $%
Q_{P}^{\prime }$ in true concurrent process calculus system $P$.

We write $(C_{M},M)\preceq _{TM}(Q_{P},R_{P})$ if there is a simulation $R$
such that $((C_{M},M),(Q_{P},R_{P}))\in R$\textbf{.}

In the following, we show that a Turing machine with configuration can be
simulated by a true concurrent process system.

\textbf{Proposition 1.} For any Turing machine $M=\langle Q,\Gamma ,b,\Sigma
,\delta ,q_{0},F\rangle $ with configuration $C_{M}$, there is a true
concurrent process system $P=(Q_{P},R_{P})$ with process $Q_{P},$ such that $%
(C_{M},M)\preceq _{TM}(Q_{P},R_{P}).$

Proof. Let $M=\langle Q,\Gamma ,b,\Sigma ,\delta ,q_{0},F\rangle $ be a
Turing machine, $C_{M}$ is a configuration of $M$, we construct $P\equiv
Trans_{T}(C_{M},M)=(Q_{P},R_{P})=((\nu L)(P_{1}\otimes _{L}P_{2}\otimes
_{L}P_{3}),R_{P}).$ Let $R=\{((C_{M},M),(Q_{P},R_{P}))$ \TEXTsymbol{\vert} $%
Trans_{T}(C_{M},M)=(Q_{P},R_{P})\}$. By the construction of $Trans_{T},$ if $%
C_{M}\longrightarrow _{M}C_{M}^{\prime },$ then $(\nu L)(P_{1}\otimes
_{L}P_{2}\otimes _{L}P_{3})\overset{\{\{\tau \}\}}{\longrightarrow }_{P}(\nu
L)(P_{1}^{\prime }\otimes _{L}P_{2}^{\prime }\otimes _{L}P_{3}^{\prime }),$
and $((C_{M}^{\prime },M),(Q_{P}^{\prime },R_{P}))\in $ $R,$ where $%
Q_{P}^{\prime }=(\nu L)(P_{1}^{\prime }\otimes _{L}P_{2}^{\prime }\otimes
_{L}P_{3}^{\prime }).$ Therefore $(C_{M},M)$ $\preceq _{TM}(Q_{P},R_{P}).$

A Petri net is a mathematical modeling language and a graphical tool used
for the specification and analysis of concurrent, asynchronous systems. A
Petri net consists of places, transitions, and arcs. Arcs run from a place
to a transition or vice versa, never between places or between transitions.
The places from which an arc runs to a transition are called the input
places of the transition; the places to which arcs run from a transition are
called the output places of the transition. Graphically, places in a Petri
net may contain a discrete number of marks called tokens. Any distribution
of tokens over the places will represent a configuration of the net called a
marking. In an abstract sense relating to a Petri net diagram, a transition
of a Petri net may fire if it is enabled, i.e. there are sufficient tokens
in all of its input places; when the transition fires, it consumes the
required input tokens, and creates tokens in its output places. A firing is
atomic, i.e. a single non-interruptible step.

\textbf{Definition 10.} A Petri net is a triple $N=(P,T,F)$ where:

(1) $P$ and $T$ are disjoint finite sets of places and transitions,
respectively.

(2) $F\subseteq (P\times T)\cup (T\times P)$ is a set of arcs (or flow
relations).

\textbf{Definition 11.} The transformation of Petri net $%
N=(P_{N},T_{N},F_{N})$ with configuration $C_{N}\subseteq P_{N}$ to true
concurrent process calculus system $Trans_{P}(C_{N},N)=P=(Q_{P},R_{P})$ as
follows.

We construct process corresponding to configuration $C_{N}$ as follows: The
process $Q_{P}=c_{1}\otimes _{\varnothing }...\otimes _{\varnothing }c_{m},$
where $\{c_{1},...,c_{m}\}$ is the set of places $C_{N}.$ The set of basic
transition rules $\Delta _{P}$ is constructed as follows: $p_{1}\otimes
_{\varnothing }...\otimes _{\varnothing }p_{m}\overset{\{\{t\}\}}{%
\longrightarrow }q_{1}\otimes _{\varnothing }...\otimes _{\varnothing }q_{n}$
if $\{(x,t)$ \TEXTsymbol{\vert} $(x,t)\in
F_{N}\}=\{(p_{1},t),...,(p_{m},t)\} $ and $\{(t,y)$ \TEXTsymbol{\vert} $%
(t,y)\in F_{N}\}=\{(q_{1},t),...,(q_{n},t)\}$ for given $t$.

The definition of simulation $\preceq _{PN}$can be given as follows.

\textbf{Definition 12.} For Petri net $N=(P_{N},T_{N},F_{N})$ with
configuration $C_{N}$ and true concurrent process calculus system $%
P=(Q_{P},R_{P}),$ a relation $R$ is a simulation if whenever $%
((C_{N},N),(Q_{P},R_{P}))\in R$, $C_{N}\overset{\{\{t_{1},...,t_{n}\}\}}{%
\longrightarrow }_{N}C_{N}^{\prime }$ implies that there exists $%
Q_{P}^{\prime }$ such that $Q_{P}\overset{\{\{t_{1},...,t_{n}\}\}}{%
\longrightarrow }_{P}Q_{P}^{\prime }$ and $((C_{N}^{\prime
},N),(Q_{P}^{\prime },R_{P}))\in R,$ where $C_{N}\overset{%
\{\{t_{1},...,t_{n}\}\}}{\longrightarrow }_{N}C_{N}^{\prime }$ denotes that
Petri net $N$ transforms from configuration $C_{N}$ to configuration $%
C_{N}^{\prime }$ by transitions $t_{1},...,t_{n},$ and $Q_{P}\overset{%
\{\{t_{1},...,t_{n}\}\}}{\longrightarrow }_{P}Q_{P}^{\prime }$ denotes that
process $Q_{P}$ performs action $\{\{t_{1},...,t_{n}\}\}$ to process $%
Q_{P}^{\prime }$ in true concurrent process calculus system $P$.

We write $(C_{N},N)\preceq _{PN}(Q_{P},R_{P})$ if there is a simulation $R$
such that $((C_{N},N),(Q_{P},R_{P}))\in R.$

In the following, we show that a Petri net with configuration can be
simulated by a true concurrent process system.

\textbf{Proposition 2.} For any Petri net $N=(P_{N},T_{N},F_{N})$ with
configuration $C_{N}$, there is a true concurrent process system $%
P=(Q_{P},R_{P})$ with initial process $Q_{P}$, such that $(C_{N},N)\preceq
_{PN}(Q_{P},R_{P}).$

Proof. Let $N=(P_{N},T_{N},F_{N})$ with initial configuration $C_{N},$ we
construct $P=Trans_{P}(C_{N},N)=(Q_{P},R_{P}).$ Let $R=%
\{((C_{N},N),(Q_{P},R_{P}))$ \TEXTsymbol{\vert} $%
Trans_{P}(C_{N},N)=(Q_{P},R_{P})\}$. By the construction of $Trans_{P},$ if $%
C_{N}\overset{\{\{t_{1},...,t_{n}\}\}}{\longrightarrow }_{N}C_{N}^{\prime },$
then $Q_{P}\overset{\{\{t_{1},...,t_{n}\}\}}{\longrightarrow }%
_{P}Q_{P}^{\prime }$ and $((C_{N}^{\prime },N),(Q_{P}^{\prime },R_{P}))\in
R. $ Therefore $(C_{N},N)$ $\preceq _{PN}(Q_{P},R_{P}).$

\section{Lattice-valued Performance Evaluation Language $PEL^{n}$}

Temporal logic is a kind of modal logic that deals with reasoning about
propositions whose truth values can change over time. It extends classical
propositional logic by introducing temporal operators that allow the
expression of time-dependent statements. In this section, we generalize
temporal logic $mu$ calculus to a complete-lattice-monoid-valued performance
evaluation language named $PEL^{n}$ as the performance specification of
systems$.$

\subsection{Syntax of $PEL^{n}$}

In the following, we give the definition of syntax of $PEL^{n}$ formulas
where $n$ denotes $n$ processes parameters of formulas.

\textbf{Definition 13. }Syntax of $PEL^{n}.$

We first give syntax of process context as follows, named $Context$:

$\Sigma ::=P$ $|$ $[X]$ $|$ $\Sigma _{1};\Sigma _{2}$ $|$ $\Sigma
_{1}+\Sigma _{2}$ $|$ $\Sigma _{1}\otimes _{L}\Sigma _{2}$ $|$ $(\nu
a)\Sigma $, where $a\in AtomAct,$ $L\subseteq AtomAct,$ $P\in PROC,$ $X$ is
a process variable, $[X]$ is a context hole$.$

In the following, we assume that process contexts are not $\alpha $%
-convertible ($\Sigma _1\equiv _\alpha \Sigma _2$, if $\Sigma _2$ can be
obtained from $\Sigma _1$ by a finite number of changes of bound names and
variables.)

Syntax of $PEL^{n}$ is given as follows.

$\Phi ::=A$ \TEXTsymbol{\vert} $F$ \TEXTsymbol{\vert} $\Phi _{1}\odot
_{k}\Phi _{2}$ \TEXTsymbol{\vert} $\langle \alpha \rangle _{i}^{\odot
_{k}}.\Phi $ $|$ $\mu _{F}.\Phi _{1}(F)\overset{def}{=}\Phi _{2}(F)$ 
\TEXTsymbol{\vert} $\nu _{F}.\Phi _{1}(F)\overset{def}{=}\Phi _{2}(F)$ 
\TEXTsymbol{\vert} $\lambda (X_{1},...,X_{n}).(\Sigma _{1}[X_{1}],...,\Sigma
_{n}[X_{n}])\triangleright \Phi ,$ where $A$ is an atomic formula, $F$ is a
formula variable, $k\in \{1,...,K\},$ $\alpha \in Act,$ $\langle \alpha
\rangle _{i}^{\odot _{k}}$ denotes the next modalities for action $\alpha $
of $i$-th component$,$ $\Sigma \lbrack X]$ is a process context$.$ We
abbreviate $\lambda (X_{1},...,X_{n}).(\Sigma _{1}[X_{1}],...,\Sigma
_{n}[X_{n}])\triangleright \Phi $ as $(\Sigma _{1},...,\Sigma
_{n})\triangleright \Phi $ if $\{X_{1},...,X_{n}\}=\varnothing .$

Every formula variable $F$ in the scope of sub formula $\mu _{F}.\Phi _{1}(F)%
\overset{def}{=}\Phi _{2}(F)$ \TEXTsymbol{\vert} $\nu _{F}.\Phi _{1}(F)%
\overset{def}{=}\Phi _{2}(F)$ is called bound variable, otherwise $F$ is
called free variable. A formula $\Phi $ is closed if there is no free
variable in $\Phi .$

A formula $\Phi $ is called well named if every fixpoint operator in $\Phi $
binds a distinct variable, and free variables are distinct from bound
variables. In the following we assume that every formula is well named.

\subsection{Semantics of $PEL^{n}$}

In the following, we give the definition of semantics of $PEL^{n}$ formulas.

The value of $PEL^{n}$ formulas is in a complete-lattice-monoid.

\textbf{Definition 14. }$(V,\bot ,\top ,\leq )$ is a complete lattice if $V$
is a nonempty set, $\leq $ is a partial order relation, and for any subset $%
S,$ there exist the greatest lower bound of $S$ and the least upper bound of 
$S.$

\textbf{Definition 15. }$(V,\bot ,\top ,(\odot _{k},e_{\odot _{k}})_{(k\in
\{1,...,K\})},\leq )$ is a complete-lattice-monoid, where $(V,\bot ,\top
,\leq )$ is a complete lattice, $(V,(\odot _{k},e_{\odot _{k}})_{(k\in
\{1,...,K\})})$ is a monoid and satisfies the following conditions:

(1) $\bot \leq x\leq \top $ for any $x\in V;$

(2) $\forall x.x\odot _{k}e_{\odot _{k}}=x;$

(3) $x\odot _{k}y=y\odot _{k}x$ for any $k\in \{1,...,K\},$ $x,y\in V;$

(4) $(x\odot _{k}y)\odot _{k}z=x\odot _{k}(y\odot _{k}z)$ for any $k\in
\{1,...,K\},$ $x,y,z\in V;$

(5) if $x\leq y,$ then $x\odot _{k}z\leq y\odot _{k}z$ for any $k\in
\{1,...,K\},$ $x,y,z\in V.$

By Definition 15, operator $\odot _{k}$ can be extended to any finite set as
follows:. $\odot _{k}\varnothing =e_{\odot _{k}},$ $\odot _{k}\{x\}=x,$ $%
\odot _{k}S$ is $x_{1}\odot _{k}x_{2}\odot _{k}...\odot _{k}x_{m}$ if $%
S=\{x_{1},x_{2},...,x_{m}\}.$

\textbf{Definition 16. }$(PV,\bot _{PV},\top _{PV},(\odot
_{PV,k},e_{PV,\odot _{k}})_{(k\in \{1,...,K\})},\leq _{PV})$ is defined as
follows, where $(V,\bot ,\top ,(\odot _{k},e_{\odot _{k}})_{(k\in
\{1,...,K\})},\leq )$ is a complete-lattice-monoid.

$\leq _{PV}$ is the partial\textbf{\ }order on the function set $PV:\times
_{i=1}^{n}PROC\rightarrow V$ that $f_{1}\leq _{PV}f_{2}$ iff $%
f_{1}(p_{1},...,p_{n})\leq f_{2}(p_{1},...,p_{n})$ for any $%
(p_{1},...,p_{n})\in \times _{i=1}^{n}PROC.$

$\bot _{PV}$ is the function that $\bot _{PV}(p_{1},...,p_{n})=\bot $ for
any $p_{1},...,p_{n},$ and $\top _{PV}$ is the function that $\top
_{PV}(p_{1},...,p_{n})=\top $ for any $(p_{1},...,p_{n})\in \times
_{i=1}^{n}PROC.$

$\odot _{PV,k}$ is the operator that $(f_{1}\odot
_{PV,k}f_{2})(p_{1},...,p_{n})=f_{1}(p_{1},...,p_{n})\odot
_{k}f_{2}(p_{1},...,p_{n})$ for any $(p_{1},...,p_{n})\in \times
_{i=1}^{n}PROC.$

$e_{PV,\odot _{k}}$ is the function that $e_{PV,\odot
_{k}}(p_{1},...,p_{n})=e_{\odot _{k}}$ for any $(p_{1},...,p_{n})\in \times
_{i=1}^{n}PROC.$

\textbf{Lemma 2}. $(PV,\bot _{PV},\top _{PV},(\odot _{PV,k},e_{PV,\odot
_{k}})_{(k\in \{1,...,K\})},\leq _{PV})$ is a complete-lattice-monoid, where 
$PV:\times _{i=1}^{n}PROC\rightarrow V$ is the function set, $(V,\bot ,\top
,(\odot _{k},e_{\odot _{k}})_{(k\in \{1,...,K\})},\leq )$ is a
complete-lattice-monoid.

Proof: Since $(V,\bot ,\top ,(\odot _{k},e_{\odot _{k}})_{(k\in
\{1,...,K\})},\leq )$ is a complete-lattice-monoid, by Definition 16, we
have that $(PV,\bot _{PV},\top _{PV},(\odot _{PV,k},e_{PV,\odot
_{k}})_{(k\in \{1,...,K\})},\leq _{PV})$ is a complete-lattice-monoid.

\textbf{Definition 17. }Suppose $(V,\bot ,\top ,(\odot _{k},e_{\odot
_{k}})_{(k\in \{1,...,K\})},\leq )$ is a complete-lattice-monoid, semantics
function $Evl(\Phi ,M,W)\ $is a function belongs to $\times
_{i=1}^{n}PROC\rightarrow V,$ where $\Phi \in PEL^{n},$ $M(A)\in \times
_{i=1}^{n}PROC\rightarrow V$ for any atomic formula $A,$ $W(F)\in \times
_{i=1}^{n}PROC\rightarrow V$ for any formula variable $F$.

The value of $Evl(\Phi ,M,W)\ $is defined as follows:

$Evl(A,M,W)::=M(A).$

$Evl(F,M,W)::=W(F).$

$Evl(\Phi _{1}\odot _{k}\Phi _{2},M,W)::=Evl(\Phi _{1},M,W)\odot
_{k}Evl(\Phi _{2},M,W),$ where $(f_{1}\odot
_{k}f_{2})(p_{1},...,p_{n})=f_{1}(p_{1},...,p_{n})\odot
_{k}f_{2}(p_{1},...,p_{n}).$

$Evl(\langle \alpha \rangle _{i}^{\odot _{k}}.\Phi ,M,W)::=f^{\odot
_{k}}(PreV^{\alpha ,i}(Evl(\Phi ,M,W))),$ where $PreV^{\alpha ,i}(S)=T,$ $%
T(p_{1},...,p_{i},...,p_{n})$ is the multiset $\{\{v$ \TEXTsymbol{\vert} $%
S(p_{1},...,q_{i},...,p_{n})=v,$ $p_{i}\overset{\alpha }{\longrightarrow }%
q_{i}\}\},$ $\wp ^{M}(V)=\{MS$ \TEXTsymbol{\vert} $MS$ is a finite multiset
of $V\},$ $f^{\odot _{k}}\in (\times _{i=1}^{n}PROC\rightarrow \wp
^{M}(V))\rightarrow (\times _{i=1}^{n}PROC\rightarrow V),$ $f^{\odot
_{k}}(T)(p_{1},...,p_{n})=\odot _{k}(T(p_{1},...,p_{n})),$ $\odot _{i}S$ is $%
s_{1}\odot _{i}s_{2}\odot _{i}...\odot _{i}s_{m}$ if $S=\{%
\{s_{1},s_{2},...,s_{m}\}\},$ $\odot _{i}\{\{\}\}=e_{\odot _{i}},$ $\odot
_{i}\{\{s\}\}=s..$

$Evl(\mu _{F}.\Phi _{1}(F)\overset{def}{=}\Phi _{2}(F),M,W)::=$ the least
solution $U$ of the equation $\rho (U)=\sigma (U)$ with respect to complete
lattice $(PV,\leq _{PV})$: $\rho (U)=Evl(\Phi _{1}(F),M,W[F\leftarrow U]),$ $%
\sigma (U)=Evl(\Phi _{2}(F),M,W[F\leftarrow U]),$ where $W[F\leftarrow U]$
is a new environment that is the same as $W$ except that $W[F\leftarrow
U](F)=U.$

$Evl(\nu _{F}.\Phi _{1}(F)\overset{def}{=}\Phi _{2}(F),M,W)::=$ the greatest
solution $U$ of the equation $\rho (U)=\sigma (U)$ with respect to complete
lattice $(PV,\leq _{PV})$: $\rho (U)=Evl(\Phi _{1}(F),M,W[F\leftarrow U]),$ $%
\sigma (U)=Evl(\Phi _{2}(F),M,W[F\leftarrow U]).$

$Evl(\lambda (X_{1},...,X_{n}).(\Sigma _{1}[X_{1}],...,\Sigma
_{n}[X_{n}])\triangleright \Phi ,M,W)\overset{def}{=}Evl(\Phi ,M,W)\circ
Cxt(\Sigma _{1}[X_{1}],...,\Sigma _{n}[X_{n}]),$ where $Cxt(\Sigma
_{1}[X_{1}],...,\Sigma _{n}[X_{n}])(P_{1},...,P_{n})=(\Sigma
_{1}[P_{1}/X_{1}],...,\Sigma _{n}[P_{n}/X_{n}]),$ $(f\circ g)(x)=f(g(x)).$

The semantics of $mu$ calculus is well-defined based on Tarski fixpoint
theorem. Similarly, we will give a generalization of Tarski fixpoint
theorem, which will guarantee the well-definedness of $PEL^{n}$ formulas.

\textbf{Proposition 3 }(Tarski fixpoint theorem)\textbf{. }Suppose $(V,\leq
) $ is a complete lattice, $f$ is an order preserving function, i.e., $x\leq
y\Longrightarrow f(x)\leq f(y),$ then $(\{x$ $|$ $x=f(x)\},\leq )$ is a
complete lattice.

\textbf{Definition 18. }$range(f)$ is the set $\{y$ \TEXTsymbol{\vert} $%
\exists x.f(x)=y\}.$

\textbf{Proposition 4. }Suppose $(V_{1},\leq _{1})$ and $(V_{2},\leq _{2})$
are complete lattices, $f:V_{1}\rightarrow V_{2}$ and $g:V_{1}\rightarrow
V_{2}$ are order preserving functions $(x\leq _{1}y\Longrightarrow f(x)\leq
_{2}f(y)$ and $x\leq _{1}y\Longrightarrow g(x)\leq _{2}g(y)),$ $%
range(g)\subseteq range(f)\ (\forall x.\exists y.g(x)=f(y)),$ $f$ is an
injective function$,$ then $(\{x$ $|$ $f(x)=g(x)\},\leq _{1})$ has the least
element.

Proof: We prove that there is the least solution $x_{\inf }$ such that $%
f(x_{\inf })=g(x_{\inf }).$ We let $f_{\inf }^{-1}(y)=\inf \{x$ \TEXTsymbol{%
\vert} $f(x)=y\}$ where $\inf S$ is the greatest lower bound of $S.$ It is
clear that $f_{\inf }^{-1}$ is an order preserving function. Since $%
range(g)\subseteq range(f),$ we have $f_{\inf }^{-1}(g(x))$ is an order
preserving function. By Tarski fixpoint theorem, there is the least fixpoint 
$z_{\inf }$ such that $z_{\inf }=f_{\inf }^{-1}(g(z_{\inf })).$ Since $%
z_{\inf }$ is a solution of $z_{\inf }=f_{\inf }^{-1}(g(z_{\inf })),$ and $f$
is an injective function$,$ we have $f(z_{\inf })=f(f_{\inf }^{-1}(g(z_{\inf
})))=g(z_{\inf }).$ Suppose $x$ is a solution of $f(x)=g(x),$ since $f$ is
an injective function$,$ we have $f_{\inf }^{-1}(g(x))=f_{\inf
}^{-1}(f(x))=x.$ $z_{\inf }$ is the least fixpoint such that $z_{\inf
}=f_{\inf }^{-1}(g(z_{\inf })).$ Therefore $z_{\inf }\leq _{1}x.$

\textbf{Proposition 5. }Suppose $(V_{1},\leq _{1})$ and $(V_{2},\leq _{2})$
are complete lattices, $f:V_{1}\rightarrow V_{2}$ and $g:V_{1}\rightarrow
V_{2}$ are order preserving functions $(x\leq _{1}y\Longrightarrow f(x)\leq
_{2}f(y)$ and $x\leq _{1}y\Longrightarrow g(x)\leq _{2}g(y)),$ $%
range(g)\subseteq range(f)\ (\forall x.\exists y.g(x)=f(y)),$ $f$ is an
injective function$,$ then $(\{x$ $|$ $f(x)=g(x)\},\leq _{1})$ has the
greatest element.

Proof: We prove that there is the greatest solution $x_{\sup }$ such that $%
f(x_{\sup })=g(x_{\sup }).$ We let $f_{\sup }^{-1}(y)=\sup \{x$ \TEXTsymbol{%
\vert} $f(x)=y\}$ where $\sup S$ is the least upper bound of $S.$ It is
clear that $f_{\sup }^{-1}$ is an order preserving function. Since $%
range(g)\subseteq range(f),$ we have $f_{\sup }^{-1}(g(x))$ is an order
preserving function. By Tarski fixpoint theorem, there is the greatest
fixpoint $z_{\sup }$ such that $z_{\sup }=f_{\sup }^{-1}(g(z_{\sup })).$
Since $z_{\sup }$ is a solution of $z_{\sup }=f_{\sup }^{-1}(g(z_{\sup })),$
and $f$ is an injective function$,$ we have $f(z_{\sup })=f(f_{\sup
}^{-1}(g(z_{\sup })))=g(z_{\sup }).$ Suppose $x$ is a solution of $%
f(x)=g(x), $ since $f$ is an injective function$,$ we have $f_{\sup
}^{-1}(g(x))=f_{\sup }^{-1}(f(x))=x.$ $z_{\sup }$ is the greatest fixpoint
such that $z_{\sup }=f_{\sup }^{-1}(g(z_{\sup })).$ Therefore $x\leq
_{1}z_{\sup }$.

The following proposition states the well-definedness of $PEL^{n}$ formulas.

\textbf{Proposition 6. }The semantics of $\mu _{F}.\Phi _{1}(F)\overset{def}{%
=}\Phi _{2}(F)$ and $\nu _{F}.\Phi _{1}(F)\overset{def}{=}\Phi _{2}(F)$ are
well defined if $Evl(\Phi _{1}(F),M,W)$ is an injective function, and $%
range(Evl(\Phi _{2}(F),M,W))\subseteq range(Evl(\Phi _{1}(F),M,W))$ for any $%
M,W$.

Proof: By Lemma 2, Proposition 4 and Proposition 5.

In the rest of this paper, we focus on the well-defined-semantics $PEL^{n}$
formulas.

\textbf{Definition 19. }The set of well-defined-semantics $PEL^{n}$ formulas
is a subset of $PEL^{n}$ formulas which satisfies the following conditions:

(1) In the case of $\Phi =A$, $F$, $\Phi _{1}\odot _{k}\Phi _{2}$, $\langle
a\rangle _{i}^{\odot _{k}}.\Phi _{1}$, $\lambda (X_{1},...,X_{n}).(\Sigma
_{1}[X_{1}],...,\Sigma _{n}[X_{n}])\triangleright \Phi _{1},$ $\Phi $ is a
well-defined-semantics $PEL^{n}$ formula if $\Phi _{1}$ and $\Phi _{2}$ are
well-defined-semantics $PEL^{n}$ formula.

(2) In the case of $\Phi =\mu _{F}.\Phi _{1}(F)\overset{def}{=}\Phi _{2}(F)$
or $\nu _{F}.\Phi _{1}(F)\overset{def}{=}\Phi _{2}(F),$ $\Phi $ is a
well-defined-semantics $PEL^{n}$ formula if $\Phi _{1}(F)$ and $\Phi _{2}(F)$
are well-defined-semantics $PEL^{n}$ formula, $Evl(\Phi _{1}(F),M,W)$ is an
injective function and $range(Evl(\Phi _{2}(F),M,W))\subseteq range(Evl(\Phi
_{1}(F),M,W))$ for any $M,W$.

Remark: If $\Phi _{1}(F)=F,$ it is easy to verify that $Evl(F,M,W)$ is an
injective function and $range(Evl(\Phi _{2}(F),M,W))\subseteq
range(Evl(F,M,W))$ for any $M,W$. Therefore $\mu _{F}.F\overset{def}{=}\Phi
_{2}(F)$ and $\nu _{F}.F\overset{def}{=}\Phi _{2}(F)$ are
well-defined-semantics $PEL^{n}$ formulas.

The following proposition provides a method to compute the least solution
and the greatest solution of equation $f(x)=g(x).$

\textbf{Proposition 7. }Suppose $(V_{1},\leq _{1})$ and $(V_{2},\leq _{2})$
are finite complete lattices, $f:V_{1}\rightarrow V_{2}$ and $%
g:V_{1}\rightarrow V_{2}$ are order preserving functions$,$ $%
range(g)\subseteq range(f),$ $f$ is an injective function$,$ then the least
element of $(\{x$ $|$ $f(x)=g(x)\},\leq _{1})$ is $\sqcup _{i\leq
|V_{1}|}(f_{\inf }^{-1}\circ g)^{i}(\bot _{1}))$ and the greatest element of 
$(\{x$ $|$ $f(x)=g(x)\},\leq _{1})$ is $\sqcap _{i\leq |V_{1}|}(f_{\sup
}^{-1}\circ g)^{i}(\top _{1}))$, where $f^{i}(x)$ denotes $i$ applications
of $f$ to $x,$ $\sqcup _{i\leq N}S_{i}=S_{0}\sqcup S_{1}\sqcup ...\sqcup
S_{N},$ $\sqcap _{i\leq N}S_{i}=S_{0}\sqcap S_{1}\sqcap ...\sqcap S_{N},$ $%
\bot _{1}$ is the least element of lattice $V_{1}$, and $\top _{1}$ is the
greatest element of lattice $V_{1},$ $|S|$ is the number of elements in $S$.

Proof: Similar to the proof of Proposition 4 and Proposition 5, we have that
the least element of $(\{x$ $|$ $f(x)=g(x)\},\leq _{1})$ is the least
element of $(\{x$ $|$ $x=f_{\inf }^{-1}(g(x))\},\leq _{1}),$ and the
greatest element of $(\{x$ $|$ $f(x)=g(x)\},\leq _{1})$ is the greatest
element of $(\{x$ $|$ $x=f_{\sup }^{-1}(g(x))\},\leq _{1})$. Since $%
(V_{1},\leq _{1})$ and $(V_{2},\leq _{2})$ are finite complete lattices, $%
f_{\inf }^{-1}(g(x))$ and $f_{\sup }^{-1}(g(x))$ are order preserving
functions, we have that the least element of $(\{x$ $|$ $x=f_{\inf
}^{-1}(g(x))\},\leq _{1})$ is $\sqcup _{i\leq |V_{1}|}(f_{\inf }^{-1}\circ
g)^{i}(\bot _{1}))$ and the greatest element of $(\{x$ $|$ $x=f_{\sup
}^{-1}(g(x))\},\leq _{1})$ is $\sqcap _{i\leq |V_{1}|}(f_{\sup }^{-1}\circ
g)^{i}(\top _{1}))$.

\textbf{Proposition 8.} If $M(A)(I_{1},...,I_{n})=M(A)(J_{1},...,J_{n})$ and 
$W(F)(I_{1},...,I_{n})=W(F)(J_{1},...,J_{n})$ for any $%
A,F,I_{1},...,I_{n},J_{1},...,J_{n}$ where $I_{1}\sim J_{1},...,I_{n}\sim
J_{n},$ then $P_{1}\sim Q_{1},...,P_{n}\sim Q_{n}$ implies $Evl(\Phi
,M,W)(P_{1},...,P_{n})=Evl(\Phi ,M,W)(Q_{1},...,Q_{n})$ for any $\Phi .$

Proof: By the induction on the structure of $\Phi .$

\subsection{Expressive Power of $PEL^{n}$}

Now we will discuss the expressive power of $PEL^{n}.$ In the case that $V=%
\mathbf{Bool},$ it is obvious that a $mu$ calculus formula is a special case
of $PEL^{n}$ formula. This means that $mu$ calculus is a sub language of $%
PEL^{n}.$ We give the $PEL^{n}$ formulas expression of some $mu$ calculus
formulas, some computation tree logic $CTL$ formulas, some dynamic logic
formulas and some Hoare logic formulas as follows.

$mu$ calculus formula $\mu X.\Phi (X)$ is expressed by $PEL^{n}$ formula $%
\mu _{X}.X\overset{def}{=}\Phi (X).$

$mu$ calculus formula $\nu X.\Phi (X)$ is expressed by $PEL^{n}$ formula $%
\nu _{X}.X\overset{def}{=}\Phi (X).$

$CTL$ formula $\mathbf{EX}\varphi $ is expressed by $PEL^{n}$ formula $\vee
_{\alpha \in Act}\langle \alpha \rangle .\varphi .$

$CTL$ formula $\mathbf{EG}\varphi $ is expressed by $PEL^{n}$ formula $\nu
_{X}.X\overset{def}{=}(\varphi \wedge \vee _{\alpha \in Act}\langle \alpha
\rangle .X).$

$CTL$ formula $\mathbf{E}\varphi _{1}\mathbf{U}\varphi _{2}$ is expressed by 
$PEL^{n}$ formula$\ \mu _{X}.X\overset{def}{=}(\varphi _{2}\vee (\varphi
_{1}\wedge \vee _{\alpha \in Act}\langle \alpha \rangle .X)).$

Dynamic logic formula $\langle P\rangle \varphi $ is expressed by $PEL^{n}$
formula$\ P\triangleright (\mu _{X}.X\overset{def}{=}(\varphi \vee (\vee
_{\alpha \in Act}\langle \alpha \rangle .X))).$

Hoare logic formula $\{\varphi _{1}\}P\{\varphi _{2}\}$ is expressed by $%
PEL^{n}$ formula$\ \varphi _{1}\rightarrow (P\triangleright (\mu _{X}.X%
\overset{def}{=}(\varphi _{2}\vee (\wedge _{\alpha \in Act}[\alpha ].X))).$

Since Hoare logic can be expressed by $PEL^{n},$ there is no complete
inference system for $PEL^{n}$ for Hoare logic has no complete inference
system.

\section{Framework of Generalized Performance Evaluation}

Performance evaluation aims to measure the performance of systems against
predefined criteria. In this section, we present a framework for generalized
performance evaluation based on the true concurrent process calculus $PROC$
and the lattice-valued performance evaluation language $PEL^{n}$.

\subsection{Problem of generalized performance evaluation}

In the following, we give the formal definition of problem of generalized
performance evaluation.

\textbf{Definition 20. }Problem of generalized performance evaluation.

Given $\Phi ,$ $M,$ $W$ and $(P_{1},...,P_{n}),$ evaluate $e$ such that $%
Evl(\Phi ,M,W)(P_{1},...,P_{n})=e,$ where $e\in V,$ $(V,\bot ,\top ,(\odot
_{k},e_{\odot _{k}})_{(k\in \{1,...,K\})},\leq )$ is a
complete-lattice-monoid, $\Phi $ is a well-defined-semantics formula of $%
PEL^{n}$.

In the following, we show that model checking, bisimulation checking and
computational complexity can be viewed as the special cases of generalized
performance evaluation problem.

(1) Model checking is a special case of generalized performance evaluation:

A $mu$ calculus model checking problem is whether $P\models \Phi ,$ where $%
\Phi \ $is a $mu$ calculus formula.$\ $This problem can be represented as a
generalized performance evaluation problem: computing $Evl(\lambda
X.[X]\triangleright \Phi ,M,W)(P)$, where value is in $V=\mathbf{Bool},$ and 
$(V,False,True,\wedge ,\vee ,\leq )$ is a complete-lattice-monoid with $%
False\leq True.$ Note that a $mu$ calculus formula is a special case of $%
PEL^{n}$ formula.

(2) Bisimulation checking is a special case of generalized performance
evaluation:

A bisimulation checking problem is whether $(P_{1},P_{2})\in \{(P_{1},P_{2})$
$|$ $\forall \alpha \in Act.\forall Q_{1}.P_{1}\overset{\alpha }{%
\longrightarrow }Q_{1},$ $\exists Q_{2}.P_{2}\overset{\alpha }{%
\longrightarrow }Q_{2},(Q_{1},Q_{2})\in F$ and $\forall \alpha \in
Act.\forall Q_{2}.P_{2}\overset{\alpha }{\longrightarrow }Q_{2},$ $\exists
Q_{1}.P_{1}\overset{\alpha }{\longrightarrow }Q_{1},$ $(Q_{1},Q_{2})\in F\}.$
This problem can be represented as a generalized performance evaluation
problem: computing $Evl((\lambda
(X_{1},X_{2}).([X_{1}],[X_{2}])\triangleright \Phi ,M,W)(P_{1},P_{2}),$
where value is in $V=\mathbf{Bool},$ $\Phi =\nu _{F}.F\overset{def}{=}\wedge
_{\alpha \in Act}([\alpha ]_{1}.\langle \alpha \rangle _{2}.F)\wedge
([\alpha ]_{2}.\langle \alpha \rangle _{1}.F)$.

(3) Computational complexity is a special case of generalized performance
evaluation:

A computational complexity problem is giving computational complexity of $P.$
This problem can be represented as a generalized performance evaluation
problem: computing $Evl(\lambda X.[X]\otimes _{\varnothing }Q\triangleright
\Phi ,M,W)(P),$ where value is in $V=\mathbf{N}\cup \{\infty
\}=\{0,1,...,\infty \},$ $(V,0,\infty ,+,\times ,\leq )$ is a
complete-lattice-monoid with $n\leq \infty $ for any $n\in \mathbf{N},$ and $%
Q$ is in the form of $P_{1}\otimes _{AtomAct}P_{2}.$ Intuitively, $Q$ is a
testing process with respect to a given complexity measure. We can construct
a procedure $P_{1}^{\prime }$ that $P_{1}^{\prime }$ records the complexity
measure value as $P_{2}$ in the form of $C_{1};C_{2};...C_{n},$ $C_{i}=C,$ $%
i=1,...,n,$ $C$ is a special process constants$.$ By Church-Turing thesis
and Proposition 1, there exists a process $P_{1}$ that simulates procedure $%
P_{1}^{\prime }.$ Therefore for a given complexity measure, we can construct
testing process $Q=P_{1}\otimes _{AtomAct}P_{2}$. $\Phi $ is defined
according to the specific complexity measure. In the case of the time
complexity, we let $\Phi =\mu _{F}.F\overset{def}{=}A\vee (\wedge _{\alpha
}\langle \alpha \rangle .F),$ where value of $\Phi $ is in $V=\mathbf{N}\cup
\{\infty \}=\{0,1,...,\infty \},$ $A$ is an atomic formula, $M(A)=n$ if $%
P_{2}=C_{1};C_{2};...C_{n},$ otherwise $M(A)=\infty $.

\subsection{Generalized Performance Evaluation Algorithm}

The generalized performance evaluation problem for $PEL^{n}$ asks the value
of $Evl(\Phi ,M,W)$ when given a well-defined-semantics $PEL^{n}$ formula $%
\Phi $, $M$, and $W.$ In general, the generalized performance evaluation
problem is incomputable.

In the following, we give a generalized performance evaluation algorithm for 
$PEL^{n}\ $in the case that the process space and the complete lattice are
both finite$.$

\quad \quad \quad \quad \quad For each $\varphi $ in $Sub(\Phi )$ do

\quad \quad \quad \quad \quad \quad \quad case $\varphi =A:Eval(\varphi
,M,W):=M(A);$

\quad \quad \quad \quad \quad \quad \quad case $\varphi =F:Eval(\varphi
,M,W):=W(F);$

\quad \quad \quad \quad \quad \quad \quad case $\varphi =\Phi _{1}\odot
_{k}\Phi _{2}:Eval(\varphi ,M,W):=Eval(\Phi _{1},M,W)\odot _{k}Eval(\Phi
_{2},M,W);$

\quad \quad \quad \quad \quad \quad \quad case $\varphi =\langle \alpha
\rangle _{i}^{\odot _{k}}.\Phi _{1}:Eval(\varphi ,M,W):=f^{\odot
_{k}}(PreV^{\alpha ,i}(Eval(\Phi _{1},M,W)));$

\quad \quad \quad \quad \quad \quad \quad case $\varphi =\mu _{F}.\Phi
_{1}(F)\overset{def}{=}\Phi _{2}(F)$ (where $\varphi $ is a
well-defined-semantics $PEL^{n}$ formula)$:$

\quad \quad \quad \quad \quad \quad \quad \quad \quad $Eval(\varphi
,M,W):=Eval(\bot _{PV},M,W),$

\quad \quad \quad \quad \quad \quad \quad \quad \quad repeat

\quad \quad \quad \quad \quad \quad \quad \quad \quad \quad \quad $\rho
:=Eval(\varphi ,M,W),$

\quad \quad \quad \quad \quad \quad \quad \quad \quad \quad \quad $%
Eval(\varphi ,M,W):=eval(\Phi _{1,\inf }^{-1}(\Phi _{2}(F)),M,W[F\leftarrow
\rho ]),$

\quad \quad \quad \quad \quad \quad \quad \quad \quad until $\rho
=Eval(\varphi ,M,W),$

\quad \quad \quad \quad \quad \quad \quad end case;

\quad \quad \quad \quad \quad \quad \quad case $\varphi =\nu _{F}.\Phi
_{1}(F)\overset{def}{=}\Phi _{2}(F)$ (where $\varphi $ is a
well-defined-semantics $PEL^{n}$ formula)$:$

\quad \quad \quad \quad \quad \quad \quad \quad \quad $Eval(\varphi
,M,W):=Eval(\top _{PV},M,W),$

\quad \quad \quad \quad \quad \quad \quad \quad \quad repeat

\quad \quad \quad \quad \quad \quad \quad \quad \quad \quad \quad $\rho
:=Eval(\varphi ,M,W),$

\quad \quad \quad \quad \quad \quad \quad \quad \quad \quad \quad $%
Eval(\varphi ,M,W):=eval(\Phi _{1,\sup }^{-1}(\Phi _{2}(F)),M,W[F\leftarrow
\rho ])$

\quad \quad \quad \quad \quad \quad \quad \quad \quad until $\rho
=Eval(\varphi ,M,W),$

\quad \quad \quad \quad \quad \quad \quad end case;

\quad \quad \quad \quad \quad \quad \quad case $\varphi =\lambda
(X_{1},...,X_{n}).(\Sigma _{1}[X_{1}],...,\Sigma _{n}[X_{n}])\triangleright
\Phi _{1}:Eval(\varphi ,M,W):=Eval(\Phi _{1},M,W)\circ Cxt(\Sigma
_{1}[X_{1}],...,\Sigma _{n}[X_{n}]).$

\quad \quad \quad \quad \quad \quad \quad return $Eval(\Phi ,M,W).$

The algorithm uses the following primitive operations:

(1) The function $Sub$, when given a formula $\Phi ,$ returns a queue of
syntactic subformulas of $\Phi $ such that if $\varphi _{1}$ is a subformula
of $\Phi $ and $\varphi _{2}$ is a subformula of $\varphi _{1},$ then $%
\varphi _{2}$ precedes $\varphi _{1}$ in the queue $Sub(\Phi ).$

(2) The function $PreV^{\alpha ,i}\in (\times _{i=1}^{n}PROC\rightarrow
V)\rightarrow (\times _{i=1}^{n}PROC\rightarrow \wp ^{M}(V)),$ where $\wp
^{M}(V)=\{MS$ \TEXTsymbol{\vert} $MS$ is a finite multiset of $V\},$ $%
PreV^{\alpha ,i}(S)=T$ and $T(p_{1},...,p_{i},...,p_{n})$ is the multiset $%
\{\{v$ \TEXTsymbol{\vert} $S(p_{1},...,q_{i},...,p_{n})=v,$ $p_{i}\overset{%
\alpha }{\longrightarrow }q_{i}\}\}.$

(3) We write $\bot _{PV}$ for the function that maps every process tuple to
the least element $\bot $ of lattice $(V,\leq ),$ and write $\top _{PV}$ for
the function that maps every process tuple to the greatest element $\top $
of lattice $(V,\leq )$.

(4) $eval(\Phi _{1,\inf }^{-1}(\Phi _{2}(F)),M,W[F\leftarrow \rho ])=\inf
\{\sigma $ \TEXTsymbol{\vert} $Eval(\Phi _{1}(F),M,W[F\leftarrow \sigma
])=Eval(\Phi _{2}(F),M,W[F\leftarrow \rho ])\},$ $eval(\Phi _{1,\sup
}^{-1}(\Phi _{2}(F)),M,W)=\sup \{\sigma $ \TEXTsymbol{\vert} $Eval(\Phi
_{1}(X),M,W[F\leftarrow \sigma ])=Eval(\Phi _{2}(F),M,W[F\leftarrow \rho
])\}.$

The partial correctness and termination of the algorithm are given in the
following.

\textbf{Proposition} \textbf{9.} Suppose that the process space and the
complete lattice are both finite, the algorithm given in the above
terminates and is correct.

Proof: Termination is guaranteed if the process space and the complete
lattice are both finite. Partial correctness of the algorithm can be proved
induction on the structure of the input formula $\Phi .$

\section{Framework of Generalized Controller Synthesis}

Controller synthesis refers to the constructing a controller for a dynamic
system to behave in a desired manner. In this section, we present a
framework for generalized controller synthesis.

\subsection{Problem of generalized controller synthesis}

In the following, we give the formal definition of problem of generalized
controller synthesis.

\textbf{Definition 21. }Problem of generalized controller synthesis.

Given $\Phi ,$ $M,$ $W$ and $e,$ find $P_{1},...,P_{i}$ for given $%
P_{i+1},...,P_{n}$ such that $Evl(\Phi ,M,W)(P_{1},...,P_{n})=e,$ where $%
e\in V,$ $(V,\bot ,\top ,\odot _{k}($where $k\in \{1,...,K\}),\leq )$ is a
complete-lattice-monoid, $\Phi $ is a well-defined-semantics formula of $%
PEL^{n}$.

In the following, we show that program synthesis, counterexample generation,
bisimilar process generation and controller synthesis can be viewed as the
special cases of generalized controller synthesis problem.

(1) Program synthesis is a special case of generalized controller synthesis:

Program synthesis focuses on the automated generation of computer programs
from specifications. The goal of program synthesis is to find a program $P$
that satisfies a given specification $\Phi $. Formally, this problem can be
viewed as solving the following problem: find a program $P,$ such that $%
Evl(\Phi ,M,W)(P)=True,$ where $\Phi \ $is a $mu$ calculus formula, and
value is in $V=\mathbf{Bool}$.$\ $Note that a $mu$ calculus formula is a
special case of $PEL^{n}$ formula.

(2) Counterexample generation is a special case of generalized controller
synthesis:

In formal verification, counterexample generation is to provide an execution
that fails to satisfy a specifications. Formally, this problem can be viewed
as solving the following problem: find a process $P,$ such that $Evl(\Phi
,M,W)(P)=False,$ where $\Phi \ $is a $mu$ calculus formula, value is in $V=%
\mathbf{Bool}$.$\ $

(3) Bisimilar process generation is a special case of generalized controller
synthesis:

Bisimilar process generation problem is to finding $P_{2}$ for given $P_{1}$
such that $P_{1}$ is bisimilar to $P_{2}$ relative to the finite set $%
MA\subseteq Act,$ $.$ This problem can be represented as a problem: find a
process $P_{2}$ for given $P_{1},$ such that $Evl(\lambda
(X_{1},X_{2}).([X_{1}],[X_{2}])\triangleright \Phi ,M,W)(P_{1},P_{2})=True,$
where value is in $V=\mathbf{Bool},$ $\Phi =\nu _{F}.F\overset{def}{=}\wedge
_{\alpha \in MA}([\alpha ]_{1}.\langle \alpha \rangle _{2}.F)\wedge ([\alpha
]_{2}.\langle \alpha \rangle _{1}.F)$.

(4) Controller synthesis is a special case of generalized controller
synthesis:

Controller synthesis is a fundamental concept in control theory and decision
theory. The goal of controller synthesis is to find a controller $P$ that
yields the given outcome value $c$ with respect to value function $\Phi $
within a specific environment $Q$. Formally, this problem can be viewed as
solving the following problem: find a process $P,$ such that $Evl(\lambda
X.[X]\otimes _{\varnothing }Q\triangleright \Phi ,M,W)(P)=c,$ where $c\in 
\mathbf{R}_{\geq 0}\cup \{\infty \},$ and $(V=\mathbf{R}_{\geq 0}\cup
\{\infty \},0,\infty ,+,\times ,\leq )$ is a complete-lattice-monoid with $%
r\leq \infty $ for any $r\in \mathbf{R}_{\geq 0}$.$\ $

\subsection{Generalized Controller Synthesis Algorithm}

The generalized controller synthesis problem for $PEL^{n}$ asks processes $%
P_{1},...,P_{i}$ for given $P_{i+1},...,P_{n}$ such that $Evl(\Phi
,M,W)(P_{1},...,P_{n})=e,$ for given a well-defined-semantics $PEL^{n}$
formula $\Phi ,$ $M,$ $W$ and $e.$ The generalized performance evaluation
algorithm for $PEL^{n}$ computes $Evl(\Phi ,M,W)$ when given a
well-defined-semantics $PEL^{n}$ formula $\Phi $, $M$, and $W.$ Let $%
Evl^{-1}(\Phi ,M,W)(e)=\{(P_{1},...,P_{n})$ \TEXTsymbol{\vert} $Evl(\Phi
,M,W)(P_{1},...,P_{n})=e\},$ where $Evl^{-1}(\Phi ,M,W)$ is a function: $%
V\rightarrow \wp (\times _{i=1}^{n}PROC).$ If $(P_{1},...,P_{n})\in
Evl^{-1}(\Phi ,M,W)(e),$ then $(P_{1},...,P_{i})$ is the solution of the
generalized controller synthesis problem for the given
well-defined-semantics $PEL^{n}$ formula $\Phi ,$ $M,$ $W$, $e$ and $%
P_{i+1},...,P_{n}.$ Therefore the generalized controller synthesis algorithm
can be reduced to the generalized performance evaluation algorithm when the
process space and the complete lattice are both finite.

\section{Conclusions}

There are many works on performance evaluation via formal methods \cite%
{BDDHP11,BHH05,HK01,JM18a,JM18b,MJ10}. The Computation Tree Measurement
Language (CTML/ASML) \cite{JM18a,JM18b,MJ10} unifies model checking and
performance evaluation over computation trees with real-valued measures; $%
PEL^{n}$ differs in three respects: (i) values range over arbitrary
complete-lattice-monoids rather than $\mathbf{R}$; (ii) $PEL^{n}${} includes
process contexts, which makes the inverse problem (synthesis) well-typed;
and (iii) $PEL^{n}$ is interpreted over a true concurrent process calculus,
so synchronization structure is explicit.

Quantitative and lattice-valued extensions of the $\mu $-calculus have been
studied extensively, for example, the quantitative $\mu $-calculus \cite%
{AFS04} which interprets fixpoint formulas over reals with min/max
operators. Proposition 4 and Proposition 5 are some adaptations of the
classical Tarski/Knaster construction to defining equations of the form $%
f(x)=g(x)$, and the well-definedness (injectivity) condition is the price of
that generalization. Compared with these works, the contribution of the
present paper is not a new fixpoint theory but the combination with process
contexts and the synthesis (inverse) problem.

For controller synthesis, beyond the continuous/hybrid frameworks of \cite%
{CK98,HSLL10,VSSLS01}, the classical discrete-event framework is supervisory
control theory~\cite{RW87}, whose fixed-point characterization of maximally
permissive supervisors we exhibited as an instance of generalized controller
synthesis in Section 5. Synthesis from temporal-logic specifications over
finite-state systems is related to our program-synthesis instance. Our
framework differs in that the specification language carries process
contexts and lattice values, but it inherits undecidability over
unconstrained (Turing-powerful) processes, so practical synthesis within the
framework must restrict the searchable process class.

One of the main aims of this paper is to present a framework of generalized
performance evaluation. We firstly propose a true concurrent process
calculus and its bisimulation. Then a complete-lattice-monoid-valued
performance evaluation language was presented. We give a framework of
generalized performance evaluation based on the true concurrent process
calculus and the complete-lattice-monoid-valued performance evaluation
language. It is shown that several problems in computer science can be
viewed as special cases of generalized performance evaluation. We also give
a generalized performance evaluation algorithm in the case that the process
space and the complete-lattice-monoid are both finite$.$ Controller
synthesis is the inverse problem of performance evaluation, which aims to
construct a controller for a dynamic system to ensure it behaves in a
desired manner \cite{CK98}, \cite{HSLL10}, and \cite{VSSLS01}. We present a
framework of generalized controller synthesis. We show several special cases
of generalized controller synthesis. An outline of generalized controller
synthesis algorithm was reduced to the generalized performance evaluation
algorithm$.$

\end{document}